\documentclass{emulateapj}
\usepackage{graphicx}
\newcommand{\asec}      {\mbox{$^{\prime \prime}  $} }

\begin{document}
\title{The effects of environment on morphological evolution between $0<z<1.2$ in the COSMOS Survey$^{\star}$}

\author{Peter Capak\altaffilmark{1}, Roberto G Abraham\altaffilmark{2}, Richard S Ellis\altaffilmark{1}, Bahram Mobasher \altaffilmark{3}, Nick Scoville\altaffilmark{1}, Kartik Sheth\altaffilmark{1,4} \& Anton Koekemoer\altaffilmark{3}}

\altaffiltext{1}{California Institute of Technology, Department of Astronomy, 105-24 Caltech, Pasadena, CA, 91125, USA}
\altaffiltext{2}{Department of Astronomy and Astrophysics, University of Toronto, 60 St. George St., Room 1403, Toronto, ON, M5S 3H8, Canada} 
\altaffiltext{3}{Space Telescope Sciences Institute, 3700 San Martin Dr., Baltimore, MD, 21218, USA}
\altaffiltext{4}{Spitzer Science Center, California Institute of Technology, 220-6 Caltech, Pasadena, CA, 91125, USA}

\altaffiltext{$\star$}{Based in part on observations with :   The NASA/ESA Hubble Space Telescope, obtained at the Space Telescope Science
Institute, which is operated by AURA Inc, under NASA contract NAS5-26555.  The Subaru Telescope, which is operated by the National Astronomical Observatory of Japan.  The MegaPrime/MegaCam, a joint project of CFHT and CEA/DAPNIA, at the Canada-France-Hawaii Telescope (CFHT) which is operated by the National Research Council (NRC) of Canada, the Institute National des Science de l'Univers of the Centre National de la Recherche and  the University of Hawaii.  The Kitt Peak National Observatory, Cerro Tololo Inter-American sObservatory and the National Optical Astronomy Observatory, which is operated by the Association of Universities for Research in Astronomy Inc. (AURA) under cooperative agreement with the National Science Foundation. }

\begin{abstract}
    We explore the evolution of the morphology density relation using the COSMOS-ACS survey and previous cluster studies.  The Gini parameter measured in a Petrosian aperture is found to be an effective way of selecting early-type galaxies free from systematic effects with redshift.  We find galaxies are transformed from late (spiral and irregular) to early (E+S0) type galaxies more rapidly in dense than sparse regions.  At a given density, the early-type fraction grows constantly with cosmic time, but the growth rate increases with density as a power law of index $0.29\pm0.02$.  However, at densities below 100 galaxies per Mpc$^{2}$ no evolution is found at $z>0.4$.   In contrast the star-formation-density relation shows strong evolution at all densities and redshifts, suggesting different physical mechanisms are responsible for the morphological and star formation transformation.  We show photometric redshifts can measure local galaxy environment, but the present results are limited by photometric redshift error to densities above $\Sigma=3$ galaxies per Mpc$^{2}$.
  \end{abstract}


\section{Introduction \label{s:intro}}

The correlation between galaxy properties such as color and morphology with galactic environment, specifically the local density of galaxies, was first noted by \citet{hubble1926}.  Later, low redshift surveys showed a high percentage of early-type (E+S0) and redder galaxies in clusters compared with the field which was quantified as the morphology-density  (T-$\Sigma$) \citep{dressler1980, dressler1997,treu2003,smith2005,postman2005} and star formation rate-density (SFR-$\Sigma$): \citep{oemler1974, melnick1977, butcher1984,  kauffmann2004} relations.  Spectroscopic redshifts from the  Sloan Digital Sky Survey (SDSS) enabled an extension of  these correlations over five orders of magnitude in $\Sigma$ at z $\leq 0.1$ \citep{goto2003,kauffmann2004}. Such studies clearly indicate the present properties of galaxies have been strongly influenced by galaxy environment, presumably by galaxy interactions.

Most investigations of galaxy evolution have focused on the star formation rates and stellar masses as determined from integrated photometry and spectroscopy \citep{cowie1995, lilly1995, dickinson2003, glazebrook2004, bauer2005, drory2005, feulner2005, juneau2005}.  In these studies star formation activity is seen to progress to less massive galaxies as the universe ages -- a phenomenon often called ``downsizing" \citep{cowie1995}. Nevertheless, the process of star formation in galaxies is complex and poorly understood \citep{somerville2005}.  On the other hand, the morphology of a galaxy is linked to the angular momentum distribution which is simpler to quantify, and hence a valuable alternative to star formation.  The presence of a disk clearly indicates a dynamically-cold stellar population that has not been significantly disturbed, while it is likely that spheroidal systems have been heated by interactions, merging or other mechanisms.

At redshifts above $z>0.3$ ground-based imaging generally has inadequate resolution for morphological surveys since the typical size of a galaxy is $\leq$0.75\asec \citep{2004ApJ...600L.107F}, so HST data is required.  The small field of view available on HST limited previous studies of the T-$\Sigma$ relation at $z>0.3$ to a few previously known galaxy clusters and field samples \citep{dressler1997,treu2003,postman2005}.  Combining these studies, \citet{smith2005} found high density and low density regions evolved differently.  They postulated different formation times were possibly responsible for the observed effect, but were limited by the look-back time and density resolution of the available data.  The COSMOS ACS data is the first HST imaging survey able to study statistically significant samples of galaxies over a range of environments and redshifts with high resolution morphologies.

In this paper, we use the COSMOS data, combined with photometric redshifts, to quantify the evolution of the T-$\Sigma$ relation as a function of redshift over the range z $= 0.3$ to $1.2$.  These measurements are combined with previous studies in the literature \citep{dressler1980, dressler1997, postman2005, smith2005} to provide a larger dynamic range in density and span in redshift than possible with COSMOS data alone.  In Section 2 we describe the imaging data and photometric redshifts  \citep{mobasher-cosmosphotz} used for this study and the selection of our sample. The morphological parameters used to characterize the galaxies are discussed and tested in Section 3.  Details of estimating density with photometric redshifts and correspondingly poor line of sight distances, are discussed in Section 4. Finally, dependence of morphology on density and redshift is presented and discussed in Sections 6 and 7 respectively. Throughout this paper we use a standard cosmology with $\Omega_v=0.7$, $\Omega_m=0.3$, and $H_o = 75$.  All literature values are converted to this cosmology unless otherwise noted.

\section{Imaging and Redshifts \label{data}}

The Hubble Space Telescope Advanced Camera for Surveys (HST-ACS) images, taken as a part of the COSMOS survey \citep{scoville-hst,koekemore-hst} are used to measure morphologies.  These images cover an area of $\sim$1.8 square degrees in $F814W$ and 81 square arc-minutes in $F475W$ with single orbit exposures.  The completeness is 50\% in F814W for a galaxy
0.5\asec in diameter with F814W$_{AB} = 26$ magnitude \citep{scoville-hst}. The point source depth is $\sim2$ mag deeper but this is not relevant to morphological studies. The median image quality is $0.05\arcsec$ and $0.08\arcsec$ (FWHM)  in F475W and F814W respectively .

Photometric redshifts are determined from multi-band ground-based photometry \citep{capak2006data} as described in \citet{mobasher-cosmosphotz}.  Our present investigation is focused on
redshifts z $= 0.3$ to $1.2$ for which the photometric redshifts have an accuracy $\sigma_z / (1+z) = 0.03$ for galaxies with I$_{AB} < 24$ mag. The redshift accuracy was determined directly by comparison of the photometric redshifts with spectroscopic redshifts for over 900 galaxies at $z<1.2$ with I$_{AB}<24$ in the COSMOS field.

A magnitude cutoff of $F814W<24$ was adopted based on experimentation in making morphological measurements. Below this magnitude, the low surface brightness also leads to  incompleteness for a typically sized $z=1$ galaxy which is 0.75\arcsec in size \citep{2004ApJ...600L.107F,scoville-hst}.

Stars are removed from the object catalog using the SExtractor \citet{1996A&AS..117..393B} CLASS\_STAR parameter measured on the ACS F814W image.  Objects with CLASS\_STAR $\geq 0.9$ are considered to be stars.   

A total of 120,187 galaxies meet our magnitude cut of $F814W<24$ and fall outside of the masked regions on the images.  Of these, 32,958 are meet the criteria to be used in our morphology-density analysis (see Section 4 and 6).  Table \ref{num-objects} gives the number of objects as a function of redshift.

\begin{deluxetable}{ccc}
\tabletypesize{\scriptsize}
\tablecaption{Number of Objects in Redshift Bins \label{num-objects}}
\tablehead{
\colhead{Redshift }&\colhead{$F814W < 24$}&\colhead{$M_{\rm V}\leq -21.2 +$}\\
\colhead{Range}    &\colhead{}	                      &\colhead{Look Back Time}
}
\startdata
$0.2<z\leq0.4$	& 18250 &	2154\\
$0.4<z\leq0.6$	& 17419 &	2817\\
$0.6<z\leq0.8$	& 27018 &	6193\\
$0.8<z\leq1.0$	& 19062 &	7829\\
$1.0<z\leq1.2$	& 13075 &	7799\\
$1.2<z\leq1.4$	& 7975   &	6166\\
\enddata
\end{deluxetable}

\section{Morphological Classification \label{morph}}

Galaxies exhibit a range of morphologies which is difficult to quantify automatically.  So, classification by eye is often used to test the efficacy of automated classifiers.  At high redshifts two additional problems arise : surface brightness dimming ($\propto (1+z)^4$)  reduces the visibility of disk identifying features such as arms and bars and redshifting (or band shifting) means that the observed visible bands increasingly sample the rest frame ultraviolet (UV). The UV is unlikely to provide a reliable classification of the major galactic stellar components since it is highly biased in favor of the youngest star forming regions and areas of low extinction.  These redshift dependent effects introduce systematics which are difficult to separate from the desired evolutionary trends.

Most of the automated morphological classification schemes \citep{abraham1994, abraham1996, CAS} have focussed on separating elliptical, disk and irregular systems; these galaxies exhibit a wide variety of structure and their appearances are very dependent on viewing angle and other projection effects. On the other hand, dynamically hot systems (E+S0 galaxies) are generally more centrally concentrated (therefore less susceptible to surface brightness dimming) and  more spherical (therefore less subject to projection effects).  Thus, many difficulties in the morphological classification discussed above are minimized by selecting just the early-type (E+S0) population.  Specifically, independent of the method, we classify galaxies with respect to whether they are or are not early-type and do not differentiate within the late type population. This binary classification is entirely adequate for studying the T-$\Sigma$ relation so long as any systematic biases do no vary with redshift or density.

\begin{figure}
\begin{center}
\includegraphics[scale=0.7]{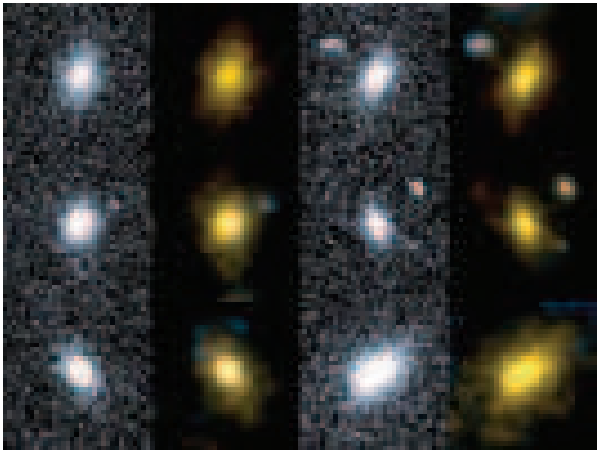}
\caption{Six galaxies with early-type visual classifications and late type automated classification are shown.  The grey scale images are the $F814W$ images used for the visual classification with a square root stretch.  The color images have the same stretch and cuts, but are made from the HST-ACS $F475W$ and $F814W$ images which were adaptively smoothed using the method described in \citet{scoville2006lss}.  Notice the low surface brightness features obvious in the color images, but hidden in the grey scale images.  \label{eyeball_bias}}
\end{center}
\end{figure}

Surface brightness has proven particularly problematic for previous morphological studies (R. Ellis \& R. Abraham private communication).  Both visual and automated classification schemes miss low surface brightness disks, resulting in systematic effects with magnitude and redshift.  These systematic effects can be mitigated by defining the edge of an object in a way which is independent of signal-to-noise per pixel.  Figure \ref{eyeball_bias} illustrates one such method for eyeball classifications.  The original HST-ACS $F814W$ images are shown together with adaptively smoothed images for several galaxies with faint disks, all of which were visually classified as early-types \citep{scoville2006lss}.  The stretch and scale factors are identical for all images.  The adaptive smoothing gives equal signal-to-noise per resolution element but with varying spatial resolution dependent on the local signal to noise ratio -- i.e. higher resolution in brighter areas.   This enhances the visibility of low surface brightness, extended features.  In the processed images, spiral structure becomes visible in all four galaxies although it was difficult to see visually in the original images.

\begin{figure}
\begin{center}
\includegraphics[scale=0.3]{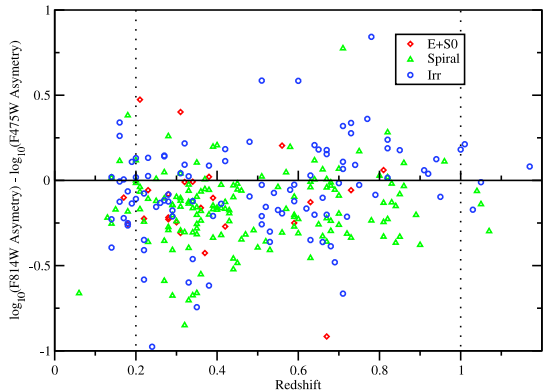}
\caption{The difference between log$_{10}$(asymetry) measured in $F475W$ and $F814W$ is shown with redshift.  Notice the systematic trends with redshift as the two bands go from sampling the rest frame optical to the rest frame UV.  Clumps of active star formation are more visible in the rest frame UV, increasing the measured asymmetry.  This transition happens at $z\simeq 0.2$ for $F475W$ and $z\simeq1.0$ for $F814W$.  These transitions are indicated with dotted lines. \label{asym-z}}
\end{center}
\end{figure}

\begin{figure}
\begin{center}
\includegraphics[scale=0.3]{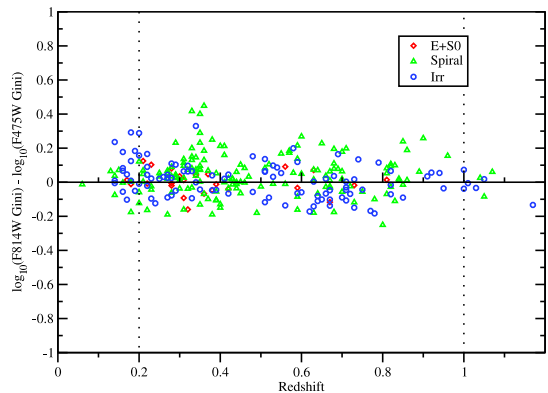}
\caption{The difference between log$_{10}$(Gini) measured in $F475W$ and $F814W$ is shown with redshift.  Notice there are no systematic trends with redshift as the two bands go from sampling the rest frame optical to the rest frame UV.  The transition from optical to UV light happens at $z\simeq 0.2$ for $F475W$ and $z\simeq1.0$ for $F814W$. These transitions are indicated with dotted lines. \label{gini-z}}
\end{center}
\end{figure}

\begin{figure}
\begin{center}
\includegraphics[scale=0.33]{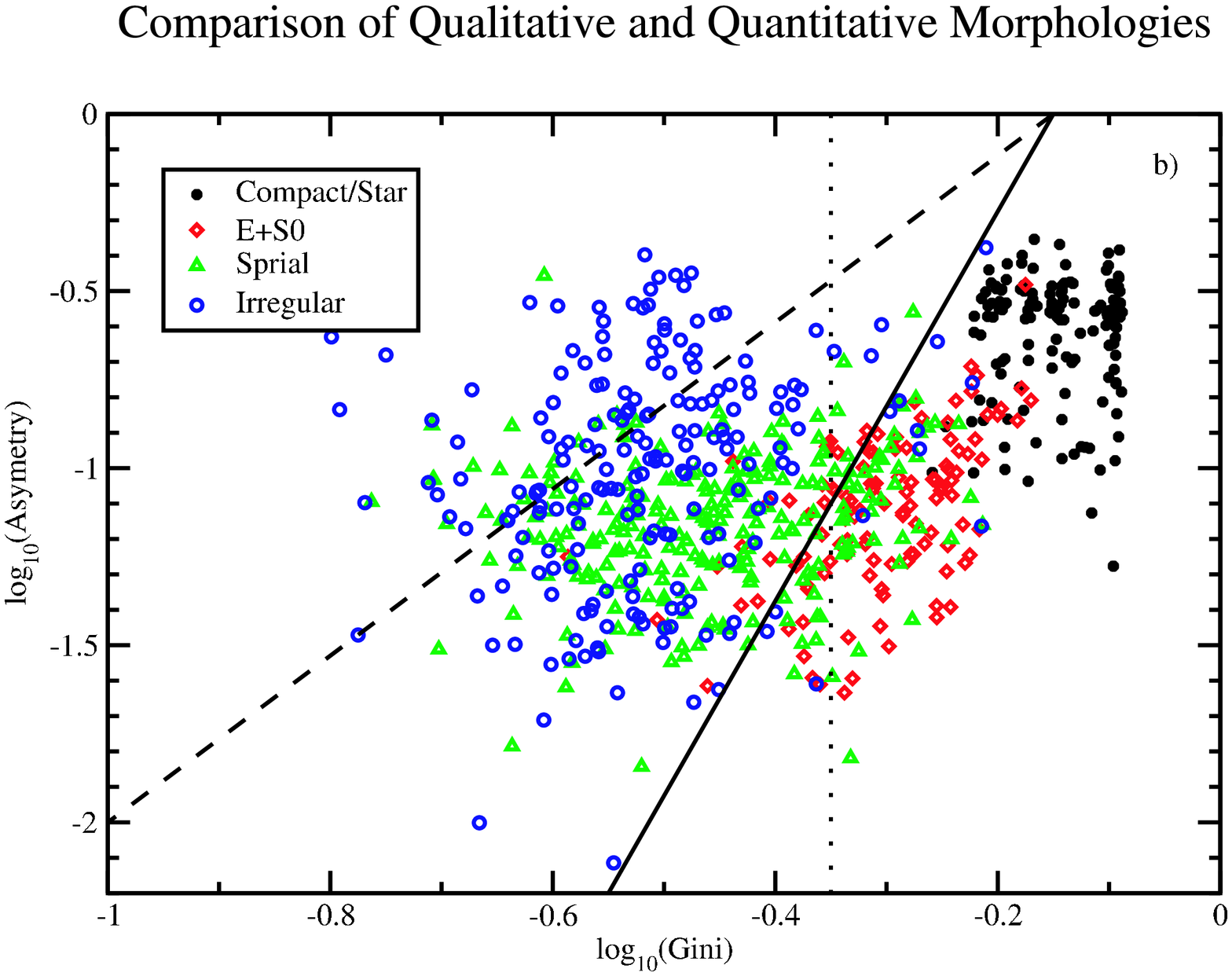}
\caption{Morphological parameters measured from the $F814W$ HST images are plotted for objects with visual morphological classification.  The Gini parameter and rotational asymmetry measured in a Petrosian aperture are plotted for objects with visual classifications.   Dividing lines are drawn between regions of predominantly irregular, spiral, and elliptical types.  The dashed line between irregular and spiral galaxies is defined as log$_{10}$(Asymmetry)$=2.353*$log$_{10}$(Gini)+0.353, while the solid line between spiral and early-type galaxies is defined as log$_{10}$(Asymmetry)$=5.500*$log$_{10}$(Gini)$+0.825$. The dotted line is a cut at log$_{10}$(Gini)=-0.35, used to select early-type galaxies in this paper. This selection gives an equivalent fraction of early-type galaxies as the Gini-Asymetry selection at $0.2 \leq z \leq 0.4$.  Stars and very compact objects appear in the upper right of this plot, but are removed in our later analysis.\label{visual_morph}}
\end{center}
\end{figure}

Petrosian apertures provide object limits (edges) independent of signal-to-noise per pixel for automated classification methods  \citep{lotz2004}.  We use a `Quasi-Petrosian' aperture to minimize the effects of surface brightness dimming.  This was constructed using a new algorithm that is intended to work for galaxies of arbitrary shape, and which has more graceful convergence properties than the usual formulation of Petrosian apertures.  (Ordinary Petrosian indices are based on circular apertures, and are not guaranteed to converge).  The details of our procedure are given in \citep{nair2006} and  only an outline  is given here. The first step is to use SExtractor to isolate the  galaxy from the sky. The flux values of the pixels in the galaxy are  then  sorted in decreasing order to construct of curve of sorted flux values.  This curve is then summed over to construct a curve of cumulative flux values, which is then multiplied by a scale factor $\eta$. The flux value where the scaled cumulative distribution intersects the sorted pixel value distribution defines a critical flux value. Only those pixels with fluxes greater than this critical value are used in calculating the `Quasi-Petrosian' coefficients. 	 
\begin{figure*}
\begin{center}
\includegraphics[scale=0.3]{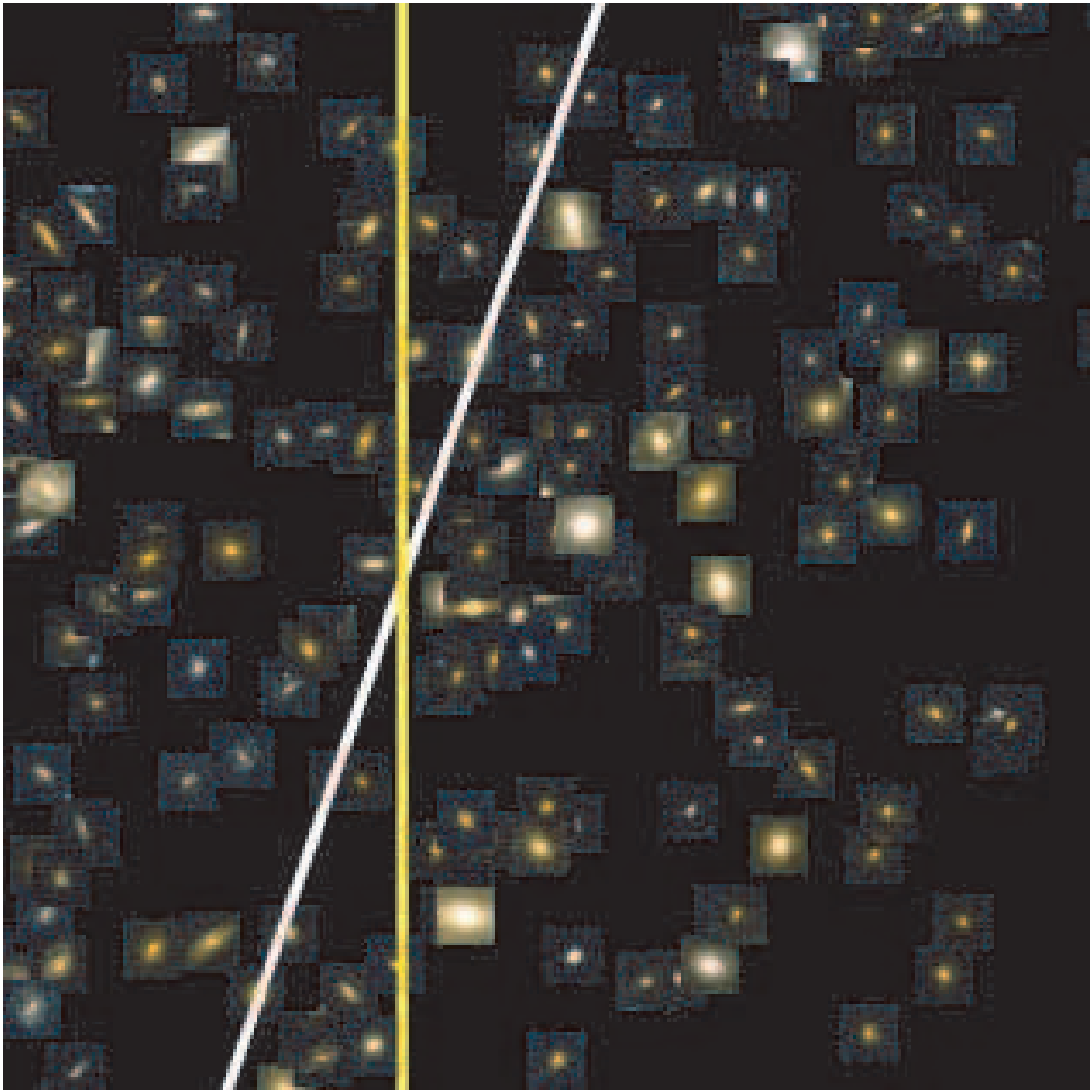}
\caption{A magnified region of Figure \ref{visual_morph} around
the division of early and late-type galaxies is shown.  The points are replaces with color images of the classified objects.  The white line corresponds to the division of early and late type galaxies in Gini and asymmetry, while the yellow line is our cut at log$_{10}$(Gini)=-0.35.  Notice the clear  division of early and late type galaxies by our cut in Gini (yellow line).  \label{gini-cut-zoom}}
\end{center}
\end{figure*}

	Rotational asymmetry and Gini are two commonly used parameters for automated morphological classification \citep{abraham1994, abraham1996, CAS}.  The asymmetry coefficient is calculated from the difference between the galaxy image and it's image rotated 180$^\circ$ about it's central peak.  The Gini parameter measures segregation of light into bright and faint pixels and is strongly correlated with the concentration of galaxy light, but the Gini parameter is a more robust indicator of galaxy morphology than the usual concentration coefficient \citep{abraham2003}.   In particular Gini is less sensitive to surface brightness effects and does not require a well defined centroid.  Specific details of our Gini and asymmetry measurements are given in \citet{2007astro.ph..1779A}. 

	The F814W HST data which covers the whole field will sample a different rest frame wavelength range at each redshift.  So any morphological classification scheme used to study galaxy evolution must be independent of the rest frame wavelength.  We can test for systematic variations due to band shifting in our data by using the the central 81 square arc minutes of COSMOS with dual band coverage (F814W and F475W).  Figure \ref{asym-z} shows the difference between asymmetry measured in F475W and F814W bands with redshift.  A shift between the classifications derived in the two bands is observed at $z\sim0.2$ where the median wavelength of the F475W band moves into the rest frame ultraviolet (UV) light while F814W still samples the rest frame visible.  A shift back to consistent classification occurs at $z\sim1.0$ where both bands sample the rest frame UV light.   As noted earlier, the rest frame UV selects regions of active star formation which tend to be irregular/clumpy; as a result, the  asymmetry is significantly higher in the rest frame UV than in the rest frame optical.  Figure \ref{gini-z} shows the difference between Gini measured in F475W and F814W bands with redshift.  The scatter is much smaller, and no systematic trends are observed with redshift.

\begin{figure*}
\begin{center}
\includegraphics[scale=0.3]{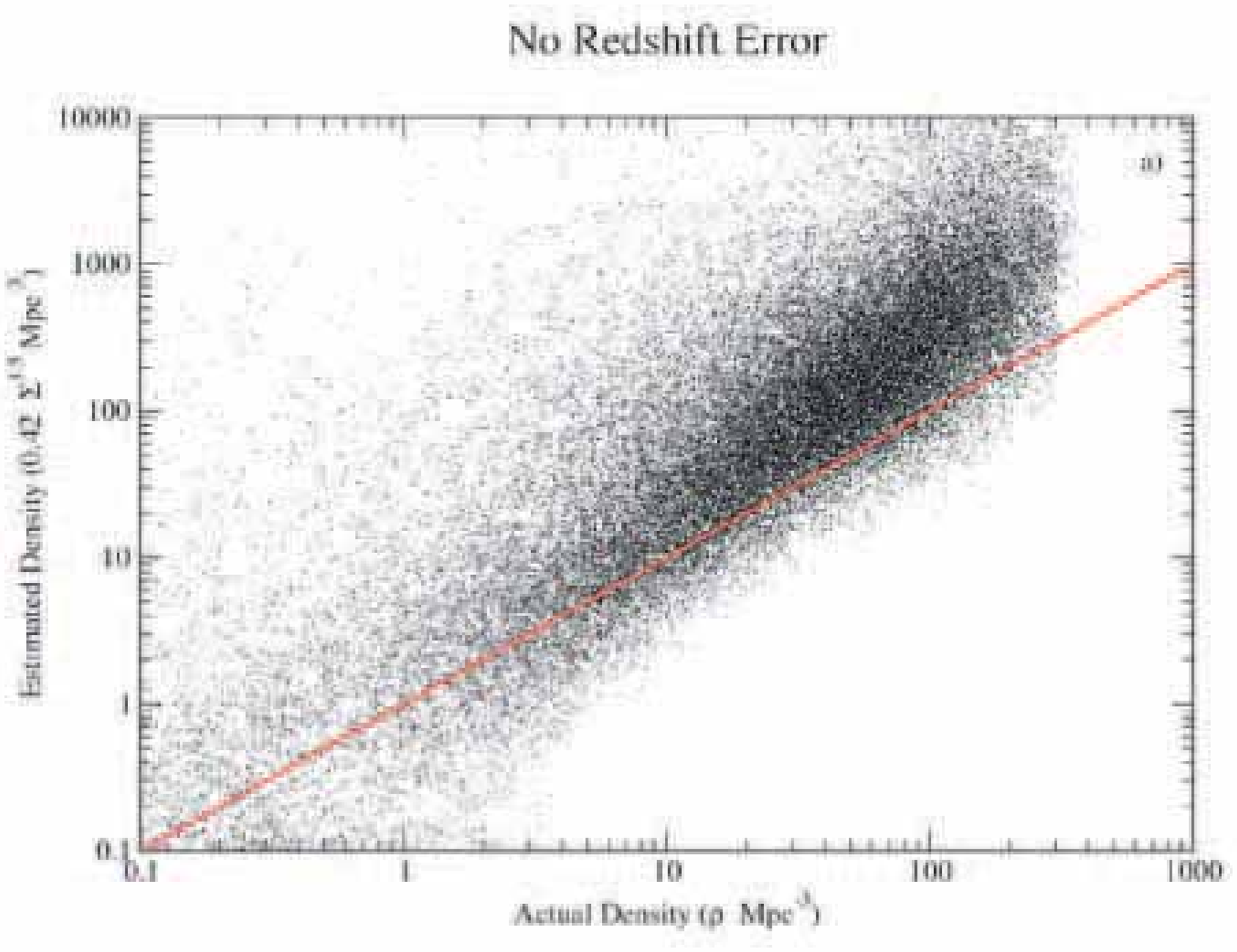}
\includegraphics[scale=0.3]{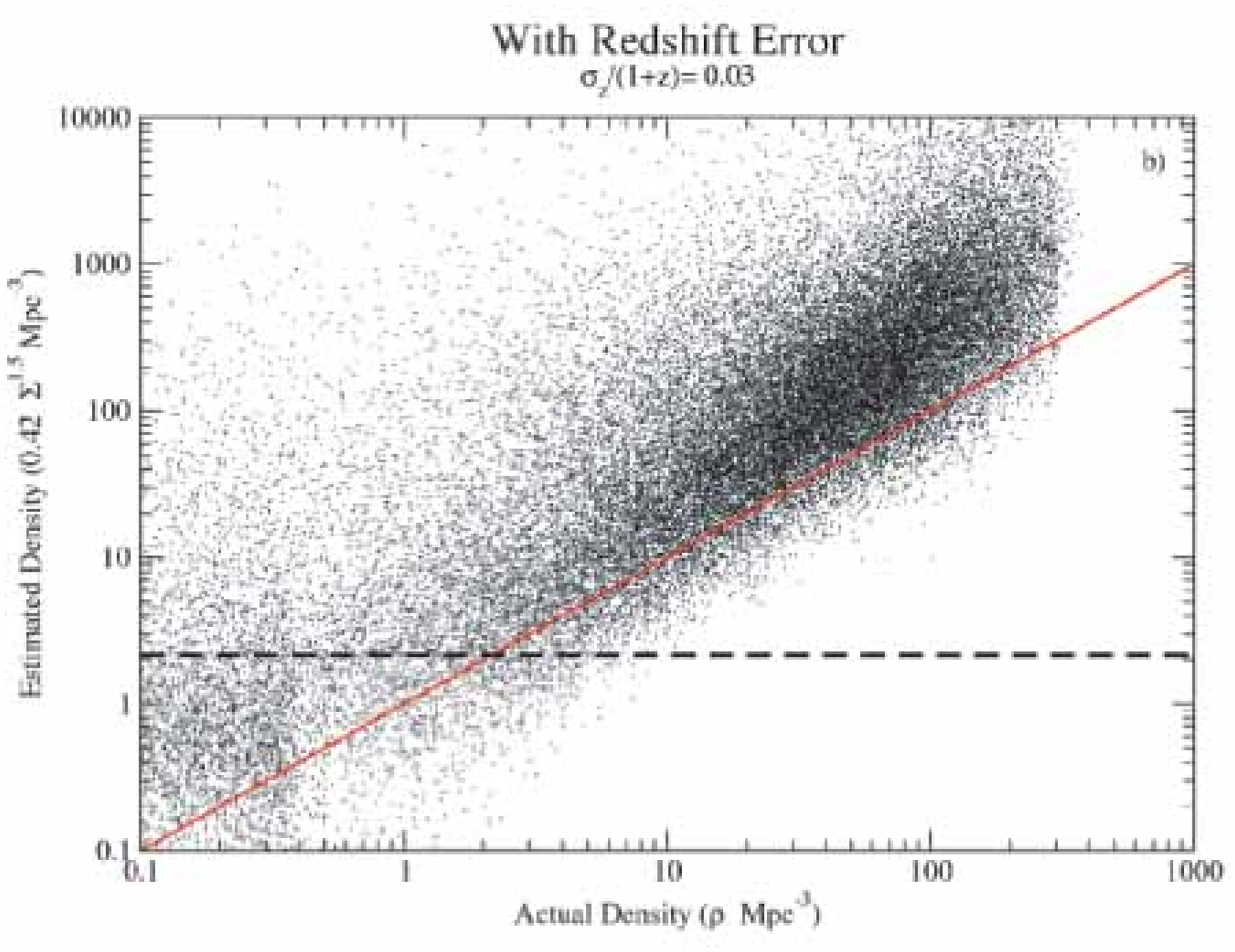}
\caption{Comparison of real-space and projected density estimates
for a simulated catalog with structures of known density.  The left
panel (a) includes no redshift error, the right panel (b) includes a
redshift error of $\sigma_z / (1+z) = 0.03$.  The dashed line in
panel b indicates the minimum measurable density, $\Sigma=3$,
estimated by Equation \ref{e:den-frac}.  A clear correlation is seen
between the actual and projected densities over several orders of
magnitude.\label{f:density-sim} }
\end{center}
\end{figure*}

	Figure \ref{visual_morph} shows a version of the \citep{abraham1996} morphology classification system compared to visual morphologies.  Visual morphologies were provided by one of us [RSE] for a complete sample of $\simeq$2000 F814W$<$22.5 galaxies in the inner, dual band coverage region using the precepts discussed in \citet{2005ApJ...625..621B}. The scatter is reasonably large due to the effects of band shifting and surface brightens dimming on the asymmetry coefficient and visual classification.   The difference between Spiral and Irregular galaxies is not important for the present investigation of the T-$\Sigma$ relation; we therefore use only the Gini parameter to classify early-type galaxies.  We chose a cut at log$_{10}$(Gini)$\leq -0.35$ as the dividing line between early and late type galaxies because it gave a similar fraction of early and late type galaxies as the \citet{abraham1996} system at $0.2\leq z \leq 0.4$.  Figure \ref{gini-cut-zoom} shows a region of Figure \ref{visual_morph} near the division line between early and late type galaxies.  The axes are identical, but the symbols have been replaced with color images of the actual galaxies.  Our specified cut in Gini coefficient clearly separates early and late type galaxies. 

\section{Density Estimate \label{s:den-estimate}}

    We adopted the N$^{th}$ neighbor projected density estimate introduced by a number of previous investigators \citep{dressler1980,dressler1997, postman2005, smith2005}.  The projected density in the vicinity of each galaxy is estimated from the distance to the N$^{th}$ neighbor.  This distance defines the radius of a circle whose area is used to estimate the surface density as: 
    
\begin{equation}
\Sigma = \frac{N}{\pi r^2} \label{e:density}
\end{equation}

{\noindent To be consistent with previous studies, we chose to use the 10$^{th}$ nearest neighbor.  A median background density is then subtracted to correct for line of sight superposition due to uncertainties in the distances (ie. redshift).  This estimator is optimal when line of sight distances are uncertain since only projected distances need to be estimated.}

    Photometric redshifts improve the nearest neighbor method in several key ways.  The rest-frame absolute luminosity of galaxies can be accurately estimated so similar galaxies (from the same part of the luminosity function) can be selected at all redshifts.  The nearest neighbor counting can be done in redshift bins rather than along the entire line of sight, reducing the background contamination and allowing multiple structures to be discriminated in the same field.  In addition, the reduction in background galaxy counts enables one to probe to much lower projected densities.

    Following \citet{smith2005} we adopt a luminosity cut of $M_{\rm V} \leq -21.2$ at $z=1$ and allow for one magnitude of passive evolution between $z=1$ and the present.  This should select a similar mix of galaxies at all redshifts.  Furthermore, maintaining a consistent magnitude cut between studies is very important because the galaxy density and morphological mix may change with magnitude \citep{benson2001}.
    
    With this magnitude cut and fading due to band shifting our F814W data are sufficiently deep at $z\leq 1.4$;  however, a redshift slice of $\pm 3 \sigma_z$ is required around each galaxy to ensure all objects in a structure are measured.  As a result, we are only able to accurately measure densities for $z\leq 1.3$.

\subsection{Accuracy of the Density Estimator \label{sig-err}}

    The accuracy of the photometric redshifts determines the minimum density to which the N$^{th}$ nearest neighbor method will work.  This minimum density can be estimated by considering a structure of density $\Sigma_s$ embedded in a  random background of density $\rho_{Bkg}$, where $\rho_{Bkg}$ is the number density of galaxies per $Mpc^2$ per redshift interval.  The projected density of background sources for a slice of thickness $\Delta z$ along the line of sight is then $\Sigma_{Bkg} = \rho_{Bkg} \Delta z$.  Assuming that $\Delta z$ is large enough to include all galaxies within the structure, the fraction of galaxies actually in the structure is given by:

\begin{equation}
F_{Real} = \frac{\Sigma_s}{\Sigma_s + \Sigma_{Bkg}} = \frac{\Sigma_s}{\rho_{Bkg} \Delta z + \Sigma_s} \label{e:den-frac}
\end{equation}

{\noindent    Equation \ref{e:den-frac} can also be inverted to define a minimum density above which a certain fraction of the galaxies will be members of a given structure. We do not need to consider Poisson error because $\Sigma$ is determined with the same number of galaxies at all densities. }

    In the present data, the largest redshift error is at $z=1.3$, where $\sigma_z=0.065$ and $\Sigma_{Bkg} = 3$Mpc$^2$.  This means 50\% of galaxies will be assigned to the correct structures at $\Sigma_s=3$Mpc$^2$  and 77\% will be correctly assigned at $\Sigma_s=10$Mpc$^2$.  At $z=0.3$, $\sigma_z=0.036$ and $\Sigma_{Bkg} = 4$Mpc$^2$, so 42\% of galaxies will be assigned to the correct structures at $\Sigma_s=3$Mpc$^2$ and 71\% will be correctly assigned at $\Sigma_s=10$Mpc$^2$.

\subsection{Relation of $\Sigma$ with Volume Density}

    Ideally, one would measure volume densities rather than projected densities.  Unfortunately, the line of sight error from photometric redshifts make volume densities difficult to measure \citep{cooper2005}.  To first order, the relation between the projected mean density, $\overline{\Sigma}$, and volume density is:

\begin{equation}
\overline{\rho} = \frac{ 3 \overline{\Sigma}}{4 r} = \sqrt{\frac{ 9\pi }{16N}}\overline{\Sigma}^{3/2} = 0.42 \overline{\Sigma}^{1.5} \label{e:rho}
\end{equation}

{\noindent if we assume spherical symmetry.}

    To see how well this relation holds we created a simulated galaxy catalog with structures of known density.  The structures had a gaussian density profile and peak densities ranging from $1$ to $300$ galaxies per Mpc$^3$, similar to the observed range in COSMOS.  The results are shown in Figure \ref{f:density-sim}a.  There is a strong correlation between the input and recovered densities for all but the sparsest regions (below $\Sigma\simeq3$ galaxies per Mpc$^2$), where our assumption of spherical symmetry breaks down.  Furthermore, the same assumption appears to overestimate the true density by a factor of $\sim 2-3$.  This is also due to our assumption of spherical symmetry, which will underestimate the true volume.  This offset is only applicable if projected densities used in this paper are converted to real space densities.  No offset is observed between projected densities measured with and without redshift error.  
    
    In addition to these effects, objects in low density regions projected in front of or behind a dense region are scattered to higher densities;  however, the fraction of these objects is less than 2\% of the total, and can be neglected.  Figure \ref{f:density-sim}b shows the effect of redshift error on the density analysis.  The scatter is larger than Figure \ref{f:density-sim}a due to the line of sight errors, however the real space density is clearly recoverable.
    
    Similar tests by \citet{cooper2005} on mock galaxy catalogs from CDM simulations agree with our results (see Table 2 and Figure 1 in \citet{cooper2005}).  However, the average density in the COSMOS data is significantly higher than those in the {Cooper} {et~al.} simulations, which leads  {Cooper} {et~al.} to conclude that the accuracy of photometric redshifts is not sufficient for the majority of galaxies.  A density of $\Sigma=10$ galaxies per Mpc$^2$ corresponds to $log_{10}($D$_5)=-0.16$ in {Cooper} {et~al.}, above which there are very few galaxies in their simulation, but a significant number of galaxies in the COSMOS data.
    
\begin{figure}
\begin{center}
\includegraphics[scale=0.33]{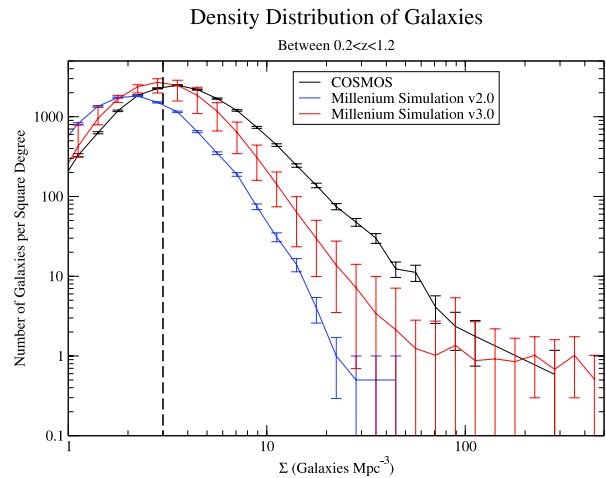}
\caption{The number of galaxies per unit area at each density are shown for COSMOS and mock catalogs from two versions of the Millennium Simulation \citep{2005Natur.435..629S}.  The vertical dashed line  indicates the minimum measurable density, $\Sigma=3$, estimated by Equation \ref{e:den-frac}.  The v2.0 mock catalog populates the dark matter haloes in a similar way to those used in \citet{cooper2005} while the v3.0 mock catalog follows galaxy orbits in detail.  The selection function, redshift error, and area of the simulations are identical to the COSMOS data.  Notice the distribution of densities in the v2.0 mock catalog is significantly lower than the actual data.  Error bars indicate the measurement Poisson error except for the v3.0 mock catalog which also includse the expected range of cosmic variance.\label{f:density-dist} }
\end{center}
\end{figure}

     To understand these differences we applied our estimator to mock galaxy catalogs provided by the Millennium simulation \citep{2005Natur.435..629S}.  These are the same simulation used by {Cooper} {et~al.},  but with dimensions and limiting magnitudes for the COSMOS survey.  As found by {Cooper} {et~al.} densities above $\Sigma = 3$ galaxies per Mpc$^2$ are successfully measured with photometric redshifts.  However, the number of galaxies at densities greater than 3 galaxies per Mpc$^2$ in the Millennium simulation is considerably lower than those in the COSMOS data.  Figure \ref{f:density-dist} shows the distribution of densities recovered from the Millennium simulation and from the COSMOS survey.  At densities above $\Sigma>3$ there are a factor of $\sim10$ more galaxies in the COSMOS survey than the v2.0 simulations.

    The discrepancy in density distribution is due to the way galaxies are distributed within dark matter haloes.  In earlier versions of the Millennium simulation galaxy orbits were not followed in detail after dark matter haloes merged (M. Kitzbichler \& S. White private communication).  This resulted in systematically fewer galaxies in high density regions.  This has been corrected in the latest versions (v3.0 and newer) of the simulation, improving the agreement, but still underestimating the number of galaxies in high density regions (See  Figure \ref{f:density-dist}).  However, \citet{mccracken-cosmos-correlation2007} find a higher amplitude in both the overall correlation function and the correlation function on small scales than predicted by the mock catalogs, which indicates the mock catalogs still tend to under-populate dense regions.

\section{Comparison With Literature Data}

\begin{deluxetable*}{lccccccc}
\tabletypesize{\scriptsize}
\tablecaption{Parameters of Previous T-$\Sigma$ Studies \label{previous-study}}
\tablehead{
\colhead{Reference }&\colhead{Redshift}&\colhead{H$_{0}$} & \colhead{$\Omega_{m}$}&\colhead{$\Omega_{\rm V}$} &\colhead{Magnitude Limit}&\colhead{Correction to} & \colhead{Correction to} \\
\colhead{}                  &\colhead{Range}  &\colhead{(km s$^{-1}$ Mpc$^{-1}$)} & \colhead{} &\colhead{} &\colhead{M$_{\rm V}\leq$ \tablenotemark{1}} & \colhead{Density\tablenotemark{2}}&\colhead{ F$_{E+S0}$}
}
\startdata
\citet{dressler1980} 	&	0.011-0.066	&	50	&	1.0	&	0.0	&	-19.75	&	-0.33 & 0.0\\
\citet{goto2003}	 	&	0.05-0.1	&	75	&	0.3	& 0.7		& -20.3\tablenotemark{3}	&	0.0 & 0.0\\
\citet{dressler1997}	&	0.37-0.56		&	50	&	1.0	&	0.0	&	-20.0	&	-0.47 & -0.045\\
\citet{treu2003}		&	0.4		&	65	&	0.3	&	0.7	&	-18.7	 &	-0.38 & -0.056 \\
\citet{smith2005}	&	0.78-1.27	&	65	&	0.3	&	0.7	&	-21.2\tablenotemark{4}	 &	-0.12 & 0.0\\
\citet{postman2005}	&	0.4-1.27	&	70	&	0.3	&	0.7	&	-20.07\tablenotemark{4} &	-0.32 & -0.045\\	
\enddata
\tablenotetext{1}{In a H$_{0}=75$, $\Omega_{m}=0.3$, $\Omega_{\rm V}=0.7$ cosmology}
\tablenotetext{2}{In units of log$_{10}(\Sigma)$.  The correction includes conversion to a H$_{0}=75$, $\Omega_{m}=0.3$, $\Omega_{\rm V}=0.7$ cosmology and an offset for the different magnitude limits.}
\tablenotetext{3}{The actual limit is M$_{\rm r}\leq-20.5$, which corresponds to M$_{\rm V}\leq-20.3$ at the median galaxy color.}
\tablenotetext{4}{The values given are for $z=1$, \citet{smith2005} allows for one magnitude of passive evolution between $z=1$ and $z=0$, while \citet{postman2005} allows for 0.8 magnitude of passive evolution.}
\end{deluxetable*}

\begin{figure}
\begin{center}
\includegraphics[scale=0.33]{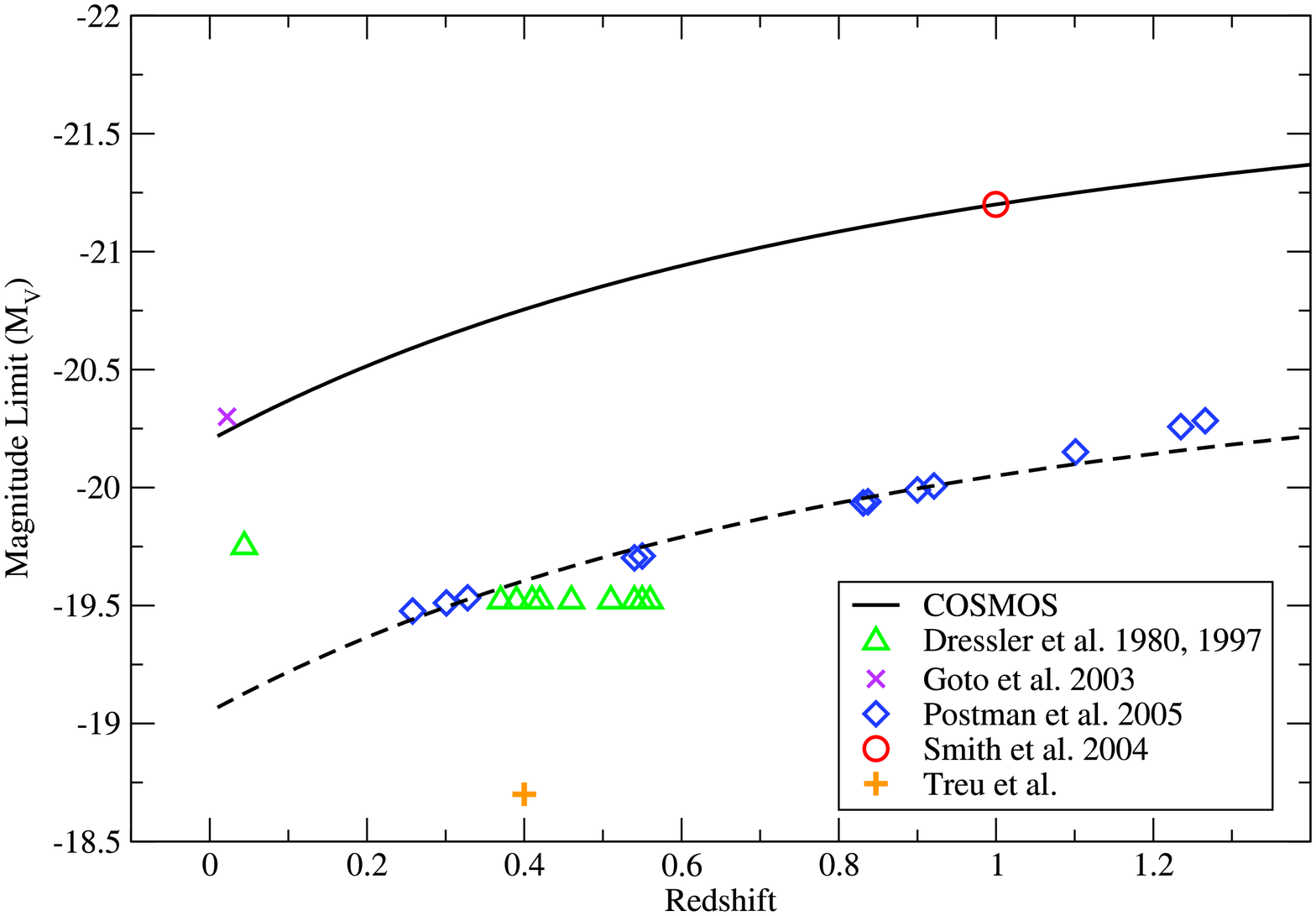}
\caption{The limiting magnitude for various studies of the T-$\Sigma$ relation is shown along with the limit used in this sample (Solid Line).  The \citet{goto2003} point has been converted from $M_{\rm r}$ to $M_{\rm V}$ assuming the median rest frame color observed in the COSMOS data between $0.2<z<0.4$ ($(V-r)=0.2$).  \citet{dressler1980,dressler1997} and \citet{postman2005} use a limiting magnitude 1 magnitude fainter than our study.  The dashed line indicates a limiting magnitude of $M_{\rm V} < -19.05$ with one magnitude of passive evolution.  \label{f:survey-limits}}
\end{center}
\end{figure}

\begin{figure}
\begin{center}
\includegraphics[scale=0.8]{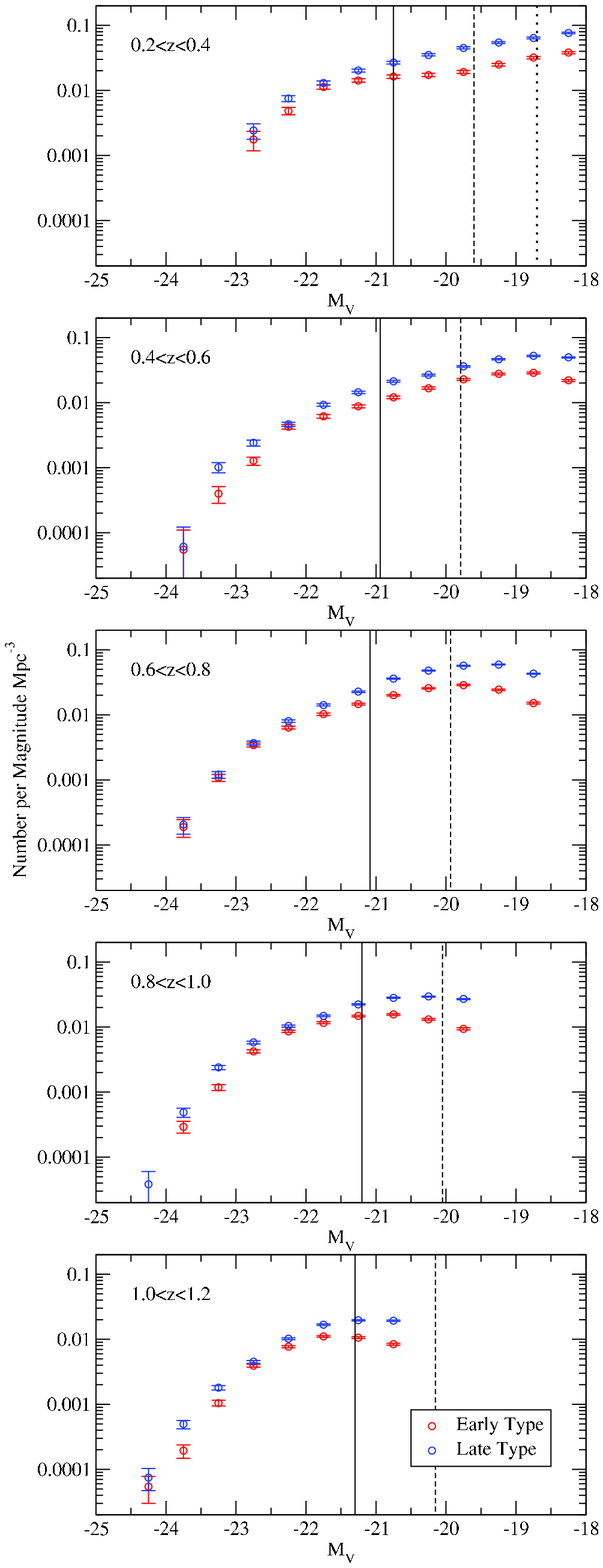}
\caption{The luminosity function of galaxies in COSMOS split by morphological type is shown for five redshift bins. These are uncorrected for incompleteness. The solid lines mark our absolute magnitude limit, the dashed line indicates the fainter limit used by \citet{dressler1980,dressler1997} and \citet{postman2005}, and the dotted line in the top panel indicates the limit used by \citet{treu2003}.  Notice the larger fraction of late type galaxies at fainter magnitudes and the fact that our data becomes incomplete at $z>0.8$ for the fainter absolute magnitude limit. \label{f:lumfunc}}
\end{center}
\end{figure}

   It is important to maintain a consistent magnitude limit when studying the morphology density relation because the galaxy density and morphological mix may change with magnitude \citep{benson2001} (see Section 4).  \citet{dressler1980,dressler1997,postman2005} and \citet{treu2003} use a magnitude limit more than a magnitude fainter than \citet{goto2003,smith2005} and our work (see Figure \ref{f:survey-limits}). To quantify the effects of the different limiting magnitudes in the literature we began by constructing morphologically selected luminosity functions uncorrected for incompleteness at the faint end.  These are shown for five redshift bins in Figure \ref{f:lumfunc}. The solid lines mark our magnitude limit, while the dashed line indicates the fainter limit used by \citet{dressler1980,dressler1997} and \citet{postman2005}.  A dotted line in the $0.2<z<0.4$ bin indicates the limit used by \citet{treu2003}.  The overall fraction of late type galaxies is clearly higher at fainter magnitudes.  In addition our data becomes incomplete at $z>0.8$ for the fainter magnitude limits.

    We estimate corrections for the differing magnitude limits by analyzing our data with three magnitude limits at $z=1$: $M_{\rm V} \leq -21.2$, $M_{\rm V} \leq -20.05$, and $M_{\rm V} \leq -19.05$.  All three analysis include 1 magnitude of passive evolution between $z=1$ and the present.  The $M_{\rm V} \leq -20.05$ limit corresponds to \citet{dressler1997} and \citet{postman2005}, while $M_{\rm V} \leq -19.05$ matches the \citet{treu2003} work at $z=0.4$.  
    
    The density measured with both fainter magnitude limits is 0.26 dex higher than that measured with $M_{\rm V} \leq -21.2$ at densities above $\Sigma>3$ galaxies per Mpc$^2$.  This offset appears to vary with density at $\Sigma<3$ galaxies per Mpc$^2$.  However, these densities are not reliable (see Section 4.1) and the trend is not apparent at higher density.  No trend with redshift was observed in either magnitude bin or at any density.
    
    The early-type fraction is 0.045 lower using a limit of $M_{\rm V} \leq -20.05$, and 0.056 lower with a limit of $M_{\rm V} \leq -19.05$ than with a limit of $M_{\rm V} \leq -21.2$.  No significant trends in these offset are  observed with density between $3<\Sigma<100$ galaxies per Mpc$^2$ or redshift between $0.2<z<0.8$.  However, we do not probe the highest densities and redshifts used in the literature. 
    
    A summary of the literature data and the correction factors applied is given in Table \ref{previous-study}.
     
\begin{figure}
\begin{center}
\includegraphics[scale=0.65]{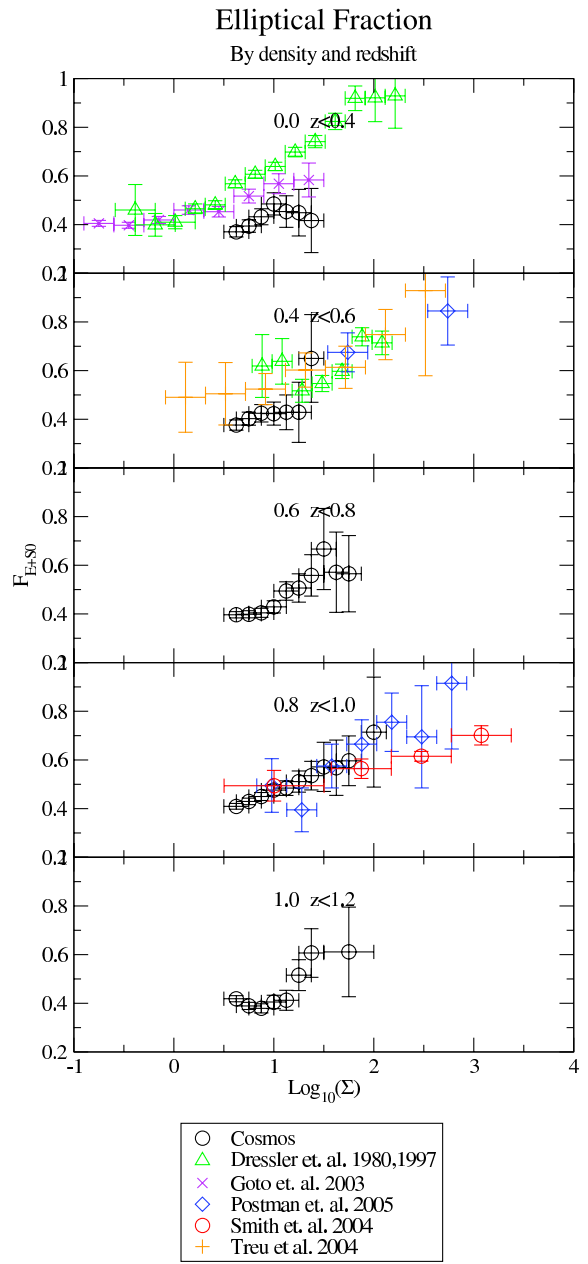}
\caption{The morphology density relation is show for five redshift intervals, $0.2\leq z < 0.4$, $0.4 \leq z < 0.6$, $0.6 \leq z < 0.8$, $0.8 \leq z < 1.0$, and $1.0 \leq z < 1.2$, along with results from similar studies.  The literature values have been converted to a cosmology with $\Omega_v=0.7$, $\Omega_m=0.3$, and $H_o = 75$ and corrected for differences in the limiting magnitude of the samples.  Note the increase in Elliptical fraction with density and decreasing redshift.  These results agree with those of \citet{dressler1980,dressler1997,goto2003,postman2005}.  The lowest density point from \citet{smith2005} is discarded due to their large redshift error (see Section 4.1). \label{morph-den-3panel}}
\end{center}
\end{figure}

\section{Results \label{s:results}}
    
    Our measurements of the T-$\Sigma$ relation, along with those taken from the literature \citep{dressler1980,dressler1997,goto2003,smith2005,postman2005}, are shown in Figure \ref{morph-den-3panel} for five redshift bins.  The densities for the literature points have been converted to a $\Omega_v=0.7$, $\Omega_m=0.3$, $H_o = 75$ cosmology and corrected for differences in the limiting magnitude.  After applying these corrections, our data is consistent with the earlier studies.  The COSMOS data significantly improves the precision of the early-type morphological fraction, and increases the redshift and density resolution, compared to earlier studies.  We find that the T-$\Sigma$ relation was already in place at $z>1$, but differs from the local relation.  As seen in previous studies \citep{postman2005, smith2005}, the early-type fraction is smaller and increases more gradually with density at $z=1$ than the at $z=0$.

\begin{figure}
\begin{center}
\includegraphics[scale=0.3]{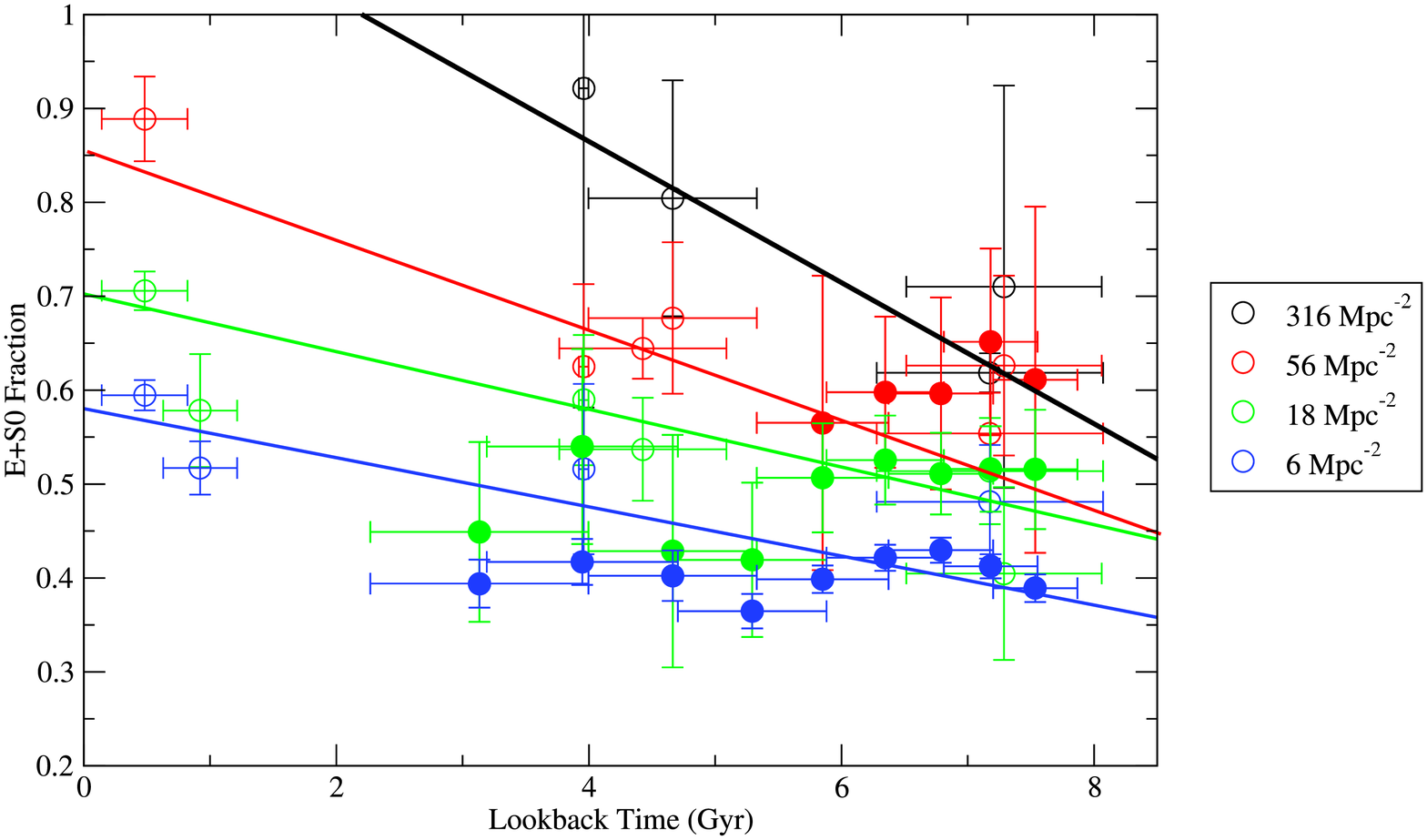}
\caption{Evolution of the elliptical fraction with redshift is shown for four densities.  Notice the increased rate of elliptical formation with increased density and the lack of evolution at $z>0.4$ in the lowest two density bins.  Open points are a compilation from \citet{dressler1980,dressler1997,treu2003,goto2003,postman2005} and \citet{smith2005}, solid points are those from this study, the lines are best fits to the data.   \label{el-frac-z}}
\end{center}
\end{figure}

\begin{figure}
\begin{center}
\includegraphics[scale=0.3]{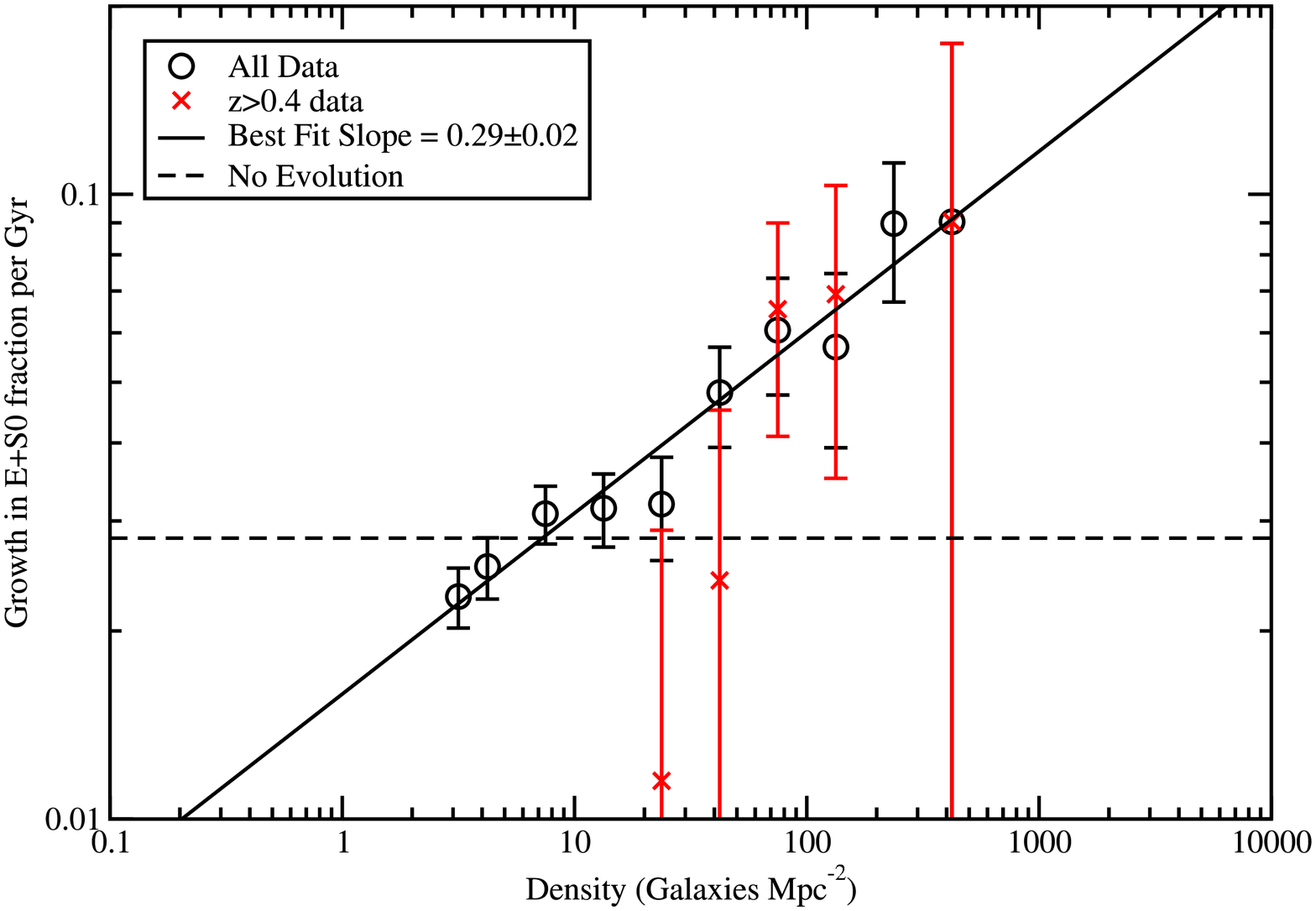}
\caption{The growth rate of the E+S0 fraction as a function of density is plotted.   The thick solid line is the best fit power law to the data with a slope of $0.29\pm0.02$, while the dashed line shows the expected trend for no change in evolution rate with density.  The red-points indicate the measured evolution if only data at $z>0.4$ is used.  No evolution is observed at densities below $\Sigma<100$ galaxies per Mpc$^2$ at these high redshifts. These rates are derived from all data presented in Figure \ref{morph-den-3panel} not the selected densities plotted in Figure \ref{el-frac-z}. \label{elip_production_rate}}
\end{center}
\end{figure}

    The evolution of the T-$\Sigma$ relation with redshift is encapsulated in Figure \ref{el-frac-z}.  This figure shows the fraction of early-type galaxies as a function of time for four different density bins---growth in the early-type fraction with look-back time is seen at all densities with more rapid growth at higher densities.  However, the majority of the evolution occurs at $z<0.4$ for densities below $\sim100$ galaxies per Mpc$^2$.  Assuming the growth rate of the early-type fraction is smooth, a line can be fit to the data at each density bin in Figure \ref{el-frac-z}.  The slope of this line yields the growth rate of the early-type fraction at any given density, which are plotted in Figure \ref{elip_production_rate}.   

\begin{figure}
\begin{center}
\includegraphics[scale=0.3]{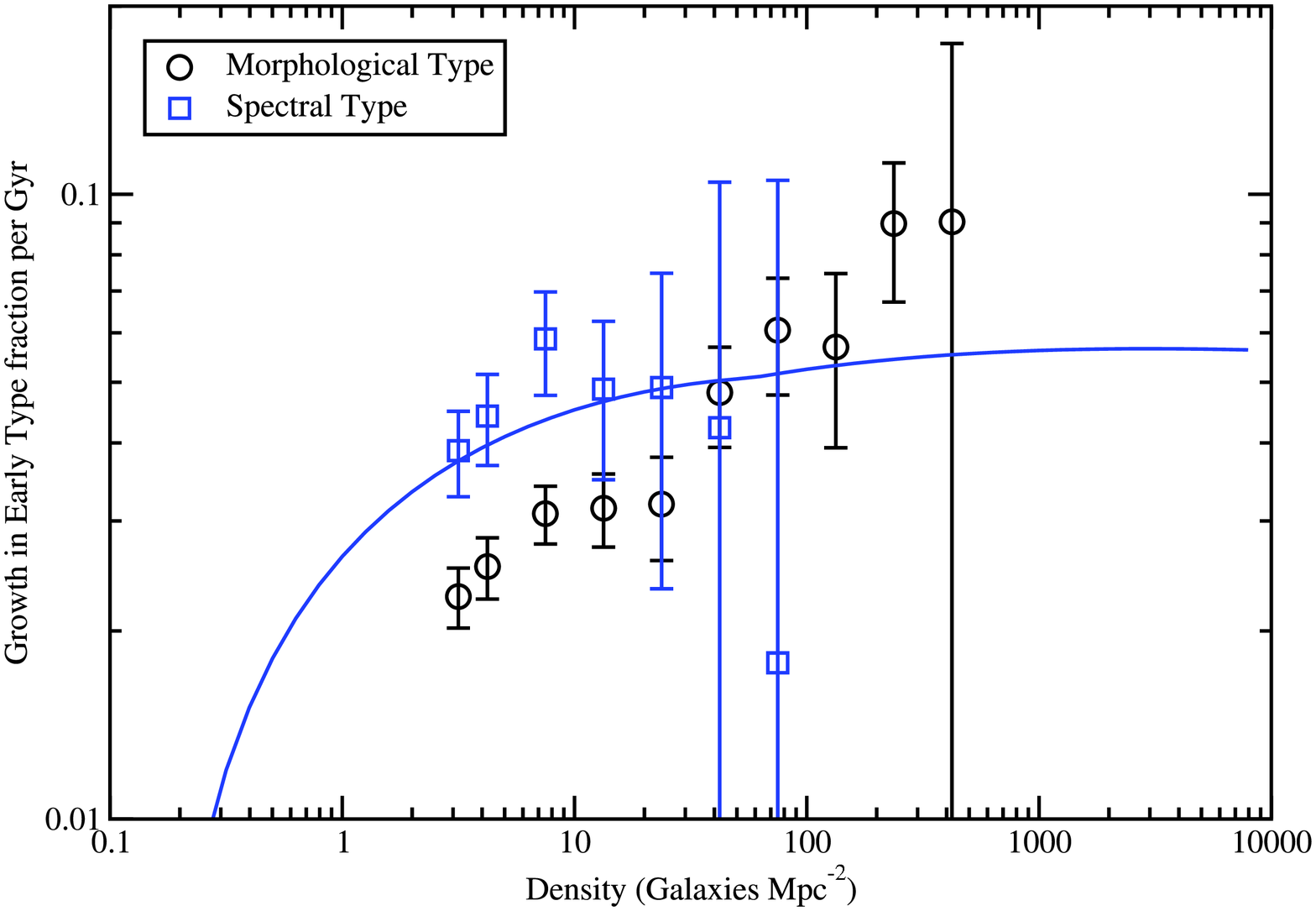}
\caption{The growth rate in the non-star-forming (passive) galaxy fraction (blue) is shown along with the growth rate in the early-type morphological fraction (black) as a function of density.   The measured rates assume the growth rate is linear with time.  The thick solid line is the linear growth rate expected by the best fit closed box model (Equations \ref{e:closed-box-cluster-model} and \ref{e:closed-box-cluster-model-rate2}) at the weighted mean lookback time of the COSMOS data (6.5Gyr).  The best fit model has $\eta=1.07 \Sigma^{0.08}$ and $t_0=-0.99*log_{10}(\Sigma) + 1.77$Gyr.  Notice the growth in the non-star-forming galaxy fraction is well fit by the closed box model, however no reasonable set of parameters can fit the growth in the morphological fraction unless more early-types are intrinsically formed in dense regions.\label{spec_production_rate}}
\end{center}
\end{figure}

	Figure \ref{elip_production_rate} shows a clear trend for more rapid early-type (E+S0) production at higher densities.  The data are well fit by a power law with a slope of $0.29\pm0.02$.  However, without the $z<0.4$ points the growth rate drops off quickly at densities below $\sim100$ galaxies per Mpc$^2$. It should be noted that points above $\Sigma\geq100$ galaxies per Mpc$^{2}$ come entirely from cluster surveys in the literature.  However, it is unlikely that the observed evolution is due to systematic effects because all three studies \citep{treu2003, smith2005, postman2005} at these redshifts and densities use the same visual morphological classification system, and the trend to stronger evolution at higher density is seen by both \citet{smith2005} and \citet{postman2005} independently.
	
	The growth in the fraction of non-star forming (passive) galaxies is shown in Figure \ref{spec_production_rate} along with the data from Figure \ref{elip_production_rate}.  The spectral types are drawn from a similar analysis in \citet{scoville2006lss}, however, the density measure and magnitude limit are the same ones used in this study, not those from \citet{scoville2006lss}.  Note the difference in the evolution measured from star formation and morphology.
    
	Many bulge dominated but elongated objects with higher asymmetry (seen near the top of Figure \ref{visual_morph}) are selected by both the {Abraham} {et~al.} and our early-type selection.  These elongated objects are likely edge on S0 galaxies.  The Gemini-Deep-Deep-Survey (GDDS) finds that many of these elongated objects have residual star formation \citep{2004AJ....127.2455A}.  Although such object with active star formation activity do not strictly meet the criteria of early-type galaxies, they are bulge dominated, and hence more dynamically relaxed than galaxies with smaller Gini co-efficent.  These objects are likely progenitors of S0's or galaxies with early-type morphology undergoing a burst of star formation.  Furthermore, since our Gini parameter is invariant with redshift, the same fraction of the galaxy population will be selected at all redshifts, implying that any contamination will not affect the observed evolution.  

	However, it is important to remember there are multiple sub-classes of early-type galaxies, and each sub-class could be evolving differently. \citet{postman2005} investigate this effect by analyzing the ellipticity distribution of cluster early-type galaxies and find no evolution in the ellipticity distribution between $z\simeq1$ and $z\simeq0.6$.  A similar analysis of our data for $\Sigma>10$ galaxies per Mpc$^2$ also finds no evolution in the ellipticity distribution of early-types.  A Kolmogorov-Smirnov test finds the distribution of ellipticities at $0.2<z<0.4$ is consistent with those at $0.8<z<1.2$ with 96.5\% certainty.  Therefore, any differential evolution in the E and S0 population is not seen in the overall ellipticity distribution of early-type galaxies.

\section{Discussion}\label{s:discussion}
	Whether the morphology-density (T-$\Sigma$) relation is an intrinsic property of cluster galaxies or a result of environmental influence has been a matter of debate for some time. In contemporary simulations of galaxy formation the majority of galaxies begin as dynamically cold, gas rich, star forming disks.  They are subsequently transformed into early-type systems through interactions with other galaxies or by losing or exhausting the gas required to form stars. In this context there are two hypothesis explaining the T-$\Sigma$ relation:  dense regions formed earlier than sparse ones, and dens regions have stronger and more frequent interactions.  A third hypothesis is that dense regions intrinsically form a higher fraction early-types, and that no transformation is needed.  
	
    	The relative importance of these hypothesis can be separated with a simple closed box model.  If galaxy clusters were closed systems one would expect the growth in the early-type fraction to slow with time as the fraction of early-type galaxies increases.  This occurs because there are fewer late type galaxies left to transform as the early-type fraction approaches 100\%.  Formally, the growth rate is given by:
	
\begin{equation}
\frac{dN_{E+S0}}{dt} = (1 - \frac{N_{E+S0}}{N})N\eta(\Sigma)
\label{e:closed-box-cluster-model-rate}
\end{equation}

{\noindent	 where the rate of change in the number density of early-type galaxies N$_{E+S0}$ is given by the total number density of galaxies N, and the conversion rate of late to early-type galaxies per unit time as a function of density, $\eta(\Sigma)$.  Integrating Equation \ref{e:closed-box-cluster-model-rate} with respect to time we get:}
 	
\begin{equation}
F_{E+S0} = 1 - (1-F_o(\Sigma))e^{-\eta(\Sigma) (t-t_0)}  
\label{e:closed-box-cluster-model}
\end{equation}

{\noindent where $F_{E+S0}$ is the early-type fraction at time t, $F_o(\Sigma)$ is the primordial fraction of early types at a given density, and $t_o$ is the cluster formation time. This equation can be differentiated to yield:}

\begin{equation}
\frac{dF_{E+S0}}{dt} = (1-F_o(\Sigma))\eta(\Sigma) e^{-\eta(\Sigma) (t-t_0)}  
\label{e:closed-box-cluster-model-rate2}
\end{equation}

	If we assume $\eta(\Sigma)=$constant and $F_o(\Sigma)=0$  the observed increase in the production rate of early-type galaxies with density suggests cluster galaxies formed significantly {\it later} than the field galaxies.  Since detailed studies of cluster ellipticals indicate they are older than their counterparts in the field \citep{bower1992, ellis1997, stanford1998,1991MNRAS.249..755L, 1996MNRAS.281..985V, 1998ApJ...493..529B, 1998AJ....116.1606P, 2000ApJ...531..137K}, we can rule out differences in formation time as the sole source of the T-$\Sigma$ relation.  A model were $\eta(\Sigma)$ increases with density, but $F_o(\Sigma)=const$, is also ruled out by the data if we require that formation time is constant or increasing with density.  A combination of earlier formation times in dense regions, increasing transformation rates with density, and an intrinsically higher early type fractions in dense regions is required to explain the observed data with a closed box model.  However, simply allowing for galaxy in-fall also explains the observed evolution.
	
	Dynamical friction or ``harassment" \citep{1958ApJ...127...17S, 1998ApJ...495..139M} is the most likely source of an increased late to early type transformation rate in dense regions.  This process operates by increasing the orbital energy of stars inside individual galaxies through tidal interactions.  The rate of momentum exchange due to dynamical friction increases with density, which naturally leads to the observed increase in transformation rate.  Furthermore, \citet{2005MNRAS.359.1415G} find differences in the velocity dispersion of cluster early and late type galaxies which can be explained if the early type galaxies underwent more dynamical friction than the late type galaxies, but can not be explained by gas stripping.  Finally, the amount of energy transfered to a galaxy through dynamical friction is proportional to the galaxy mass \citep{1958ApJ...127...17S, 1998ApJ...495..139M, 2005MNRAS.359.1415G}, so the color-magnitude and fundamental plain relations may also be a consequence.	
	
		Intrinsic differences in the galaxy formation process between clusters and the field also appear to be important in determining galaxy morphology.  Using only our data, no evolution is measured in the T-$\Sigma$ relation, but COSMOS only covers redshifts greater than 0.4 and densities below $\Sigma<100$ galaxies per Mpc$^2$.  At higher densities, evolution is independently observed by both \citet{smith2005} and \citep{postman2005}.  Nevertheless, the lack of evolution at low densities suggests the galaxy formation process may be intrinsically different in dense regions than sparse.
	
	\citet{2006ApJ...651..120B} also find intrinsic differences in the galaxy formation process may be more important than environment.  They show the star formation properties of galaxies are more correlated with galaxy mass than environment.  Environment is only important when the local galaxy density was significantly greater than the field density.  If a similar statement can be made for morphologies,  it would explain why evolution is only observed in the highest density regions.
	
	However, morphology and star formation appear to be affected by different processes.  The fraction of galaxies with early-type morphologies and the fraction with low star-formation rates evolve in different ways with density (see Figure \ref{spec_production_rate}).  These different rates of evolution can be explained if the growth in the early type fraction were driven by interactions, while the reduction in star formation was caused by gas stripping . 
	
	Assuming gas removal due to cluster interactions is responsible for truncating the star formation, we can recycle our closed box model to describe the process.  For the SFR-$\Sigma$ relation $F_o(\Sigma)=0$ because stars must form to make galaxies.  If the gas density traces the galaxy density, $\eta(\Sigma)$ becomes the rate at which star-formation is truncated and $t_o$ becomes the formation time for the cluster.  This simple model accurately reproduces the observed shape and evolution of the SFR-$\Sigma$ relation while also predicting reasonable galaxy formation times.  However, points at much lower density than what we can probe with photometric redshifts are required to confirm the gas striping hypothesis.
			
	Nevertheless, several studies of the local universe also suggest the SFR-$\Sigma$ relation is driven by gas stripping, while the T-$\Sigma$ relation is due to galaxy interactions.  \citet{2005ApJ...629..143B} find color is a better predictor of density than morphology, indicating star formation is strongly affected by environment while early-type morphologies are not necessarily the result of environment.  Further evidence is provided by \citet{2006astro.ph.11361Q}, who find both morphology and color are correlated with distance from cluster centers, but the correlation between morphology and star-formation is asymmetric.  Specifically, for a given star formation rate, the fraction of morphological early-type galaxies does not change with distance from the cluster, but at a given morphological type, the average star formation rate increases with distance from the cluster.  In addition, both \citet{dressler1997} and \citet{treu2003} show morphology is more strongly correlated with local density than distance to the cluster center, indicating the number of neighbors and hence the number of interactions is the more important quantity in determining morphology.
	
	Galaxy in-fall may also play a role in shaping the T-$\Sigma$ relation.  It is predicted by CDM models, but it is difficult to explain the small scatter in the colors and fundamental plane of cluster ellipticals if most early-type galaxies were recently transformed from field galaxies.  Nevertheless, \citet{treu2003} point out that the dynamics of galaxy clusters would erase the T-$\Sigma$ relation if the orbits of individual galaxies were not confined to regions of nearly constant density.   So, if in-fall is responsible galaxies are likely accreted smoothly via dynamical friction from adjacent regions of already elevated density.
						  
	There is no quantitative study of how harassment or gas starvation should affect the evolution of the T-$\Sigma$ relation.  \citet{benson2001} attempts to model evolution of the T-$\Sigma$ relation at low densities in cold dark matter (CDM) simulations via halo mergers.  The amplitude of the predicted evolution is correct at $\Sigma=10$ galaxies per Mpc$^2$, and no differential evolution is seen at lower densities.  Unfortunately, at higher densities, where we have the bulk of our data, the Benson et al. model is not valid because individual galaxy orbits are not tracked once the dark matter halos merge, so no comparison can be made.  
		
	Clearly more detailed models are needed, but care must be taken when comparing the results.   A magnitude-morphology relation is predicted by most CDM models, so changing the magnitude limit used to measure density will change the observed T-$\Sigma$ relation and its observed evolution.  As a result, it is very important to match the absolute magnitude cuts when comparing models and data.  Furthermore, the present measurements do not distinguish between moderate density regions on the outskirts of massive clusters and the centers of moderate mass groups.  \citet{treu2003} find this distinction is not important, nevertheless, it remains a potential source of systematic uncertainty in our measurements.
	
	In considering Figure  \ref{el-frac-z}, several limitations of our present data become evident.  Our measurement of the early-type growth rate strongly depend on the local measurements from \citet{goto2003} and \citet{dressler1980} which use a different morphological classification scheme.  Furthermore, the large gap in data between $0.1<z<0.3$, which is nearly a quarter of the cosmic time we probe, limits our ability to measure how the growth in the early-type fraction changes with time.  COSMOS has too small an area and the SDSS has neither the depth nor the spatial resolution to probe the T-$\Sigma$ relation in this gap. To probe a volume at $0.1<z<0.3$ comparable to COSMOS at $z=0.5$, a survey would have to cover $\approx 15$ square degrees with $0.6\asec$ seeing in one band.  

\section{Conclusions}
    We have developed a technique for measuring the early-type galaxy fraction with density and redshift.   The Gini parameter is found to reliably select early-type galaxies between $0.3<z<1.2$ using only the ACS F814W filter.  This selection is free from systematic effects in red-shift (band shifting) or surface brightness. 
    
    We find densities are measurable with photometric redshifts if projected densities are used.  Nevertheless, for photometric redshift accuracies of $\Delta_z / (1+z) = 0.03$ projected densities below $\Sigma=3$ galaxies per Mpc$^2$ are difficult to measure.  The observed number of galaxies at a $\Sigma>10$ galaxies per Mpc$^2$ is significantly higher than that found in earlier CDM galaxy simulations because these simulations did not follow the orbits of galaxies within dark matter haloes.  The latest simulations now follow galaxies, improving the agreement, but still underestimating the number of galaxies in dense regions.

    Using these techniques we measure the evolution of the T-$\Sigma$ relation and the growth rate of the early-type fraction with cosmic time.  We find the growth rate of the early-type fraction is increasing with density, which can not be explained by a closed box model with early formation times.  We conclude some density dependent process combined with galaxy in fall is responsible for the observed relation and evolution, with dynamical friction and harassment being the most likely mechanisms.  
    
    The SFR-$\Sigma$ relation appears to result from different processes than the T-$\Sigma$ relation.  In particular, SFR-$\Sigma$ relation evolves differently from the T-$\Sigma$ relation and there does not appear to be a direct relationship between the two.  Gas stripping is the most likely source of the SFR-$\Sigma$ relation.
    
    At $\Sigma=10$ galaxies per Mpc$^2$, the rate of dark halo interactions in CDM models predict the correct amplitude of growth in the early-type fraction, but these models are not valid at higher densities.  No current model attempts to predict our observed evolution at $\Sigma>10$, so we can only speculate as to what physical mechanisms must be at play.  Furthermore, the current generation of CDM models appear to under-estimate the number of galaxies in high density regions.

{\bf Acknowledgments} \acknowledgments {We would like to thank;
Mauro Giavalisco, Jin Koda, and Mara Salvato for their excellent
suggestions, feedback and proof reading; Marc Postman and Graham Smith
for providing early pre-prints of their papers and feedback on this work; Tomotsugu Goto and
Andrew Benson for providing data from their figures; and the
contributions of the COSMOS team.
\url{http://www.astro.caltech.edu/cosmos/}  Support for this work
was provided by NASA grant HST-GO-09822 and NSF grant OISE-0456439.
}

\bibliography{ms}
\end{document}